

\input harvmac

\Title{\vbox{\baselineskip12pt\hbox{RU-92-14}\hbox{}}}
{\vbox{\centerline{Classical and Quantum Production of Cornucopions}
\centerline{At Energies Below $10^{18}$ GeV}}}

\centerline{T. Banks$^1$ and M.
O'Loughlin$^2$}
\baselineskip18pt
\centerline{Dept. of Physics and Astronomy}
\centerline{Rutgers University}
\centerline{Piscataway, NJ 08855-0849}
\smallskip
\baselineskip18pt
\noindent
We argue that the paradoxes associated with infinitely degenerate
states, which plague relic particle scenarios for the endpoint of black
hole evaporation, may be absent when the relics are horned particles.  Most
of our arguments are based on simple observations about the classical
geometry of extremal dilaton black holes, but at a crucial point we are
forced to speculate about classical solutions to string theory in which
the infinite coupling singularity of the extremal dilaton solution is
shielded by a condensate of massless modes propagating in its infinite
horn. We use the nonsingular $c=1$ solution of (1+1) dimensional string
theory as a crude model for the properties of the condensate.  We also
present a brief discussion of more general relic scenarios based on large
relics of low mass.
\vskip1in
\noindent

\footnote{}{$^1$ (banks@physics.rutgers.edu)}
\footnote{}{$^2$ (ologhlin@physics.rutgers.edu)}
\Date{June 1992}

In a previous paper in collaboration with two of our
colleagues\ref\horned{T.Banks,A.Dabholkar,M.R.Douglas,M.O'Loughlin, {\it
Phys. Rev.}{\bf D45},(1992),3607.},
we proposed a novel solution to the puzzles of
Hawking evaporation of black holes.  Our work was based on the seminal
paper of Callan, Giddings, Harvey and Strominger (CGHS)\ref\cghs{
C.Callan,S.Giddings,J.Harvey,A.Strominger,\ {\it Phys. Rev.}{\bf
D45},(1992), 1005. },
which was in turn inspired by a number of papers on charged black holes
in dilaton gravity and string theory\ref\gibbons{G.Gibbons,S.Maeda,\ {\it
Nucl. Phys.}{\bf B298},(1988),744;\
S.Garfinkle,G.Horowitz,\hfil\break A.Strominger,\ {\it Phys. Rev.}{\bf
D43},(1991),3140.}.
The essential new
conceptual idea in all of these papers was the observation that many of
the charged black hole solutions in these theories had a geometric
structure quite different from that of the Schwarzchild black hole of
general relativity.

In a theory involving both a metric and one or more scalar fields, one
is at liberty to make Brans-Dicke transformations\ref\brans{C. Brans, R.
Dicke, {\bf Phys. Rev. 124}, 925, (1961); R. Dicke, {\bf Phys. Rev. 125},
2163, (1962).} in which
the metric is Weyl transformed by some positive function of the
scalars. From the point of view of lagrangian mechanics this is a point
transformation and no physics can depend upon it\foot{However, as we
pointed out in\ref\mink{T.Banks,M.O'Loughlin,{\it Nucl. Phys.}{\bf
B362},(1991), 649}, one must be careful to insist that the
transformation be single valued and preserve positivity of the conformal
factor.
Some recent
work in 1+1 dimensional gravity ignores these restrictions.  We believe
that such transformations change the physics of the original model in an
essential way if they are used as more than a mathematical trick to
obtain exact solutions
to the classical equations of motion.} but the geometry of
spacetime can change radically under such transformations.  Consequently,
the physics may be more transparent in one Brans-
Dicke frame rather than another.  In particular, when dealing with
effective low energy lagrangians derived from string theory, a natural
BD frame is picked out by choosing the metric along whose geodesics
strings propagate.  This is the {\it $\sigma$-model metric} used by
Garfinkle, Horowitz and Strominger\gibbons . With this choice, the
spatial geometry outside the horizon of a charged black hole is shown in
Figure 1.  Typically, there will be a large region (which we call the
horn of the black hole) with the geometry of
$I \times S^2$, where I is a real interval and $S^2$ is the round two sphere.
Most of the degrees of freedom of string theory propagate as massive
particles in this region, with only a few massless two dimensional
fields.  In particular, the extremal charged black hole has a completely
static metric, with no horizon, no singularity, and an infinite horn.

It was suggested in the paper of CGHS
and explained in our previous paper\horned that the novel geometry
of dilaton black holes could provide an intuitive resolution of most of
the puzzles of Hawking evaporation.  Although the black hole appears to
an outside observer to be a rather small object, confined to a bounded
region in space, in reality its horn contains a potentially infinite
volume, which can serve as a repository of information and conserved
quantum numbers.  In particular, Bekenstein's upper
bound\ref\bekenstein{J.D.Bekenstein,{\it Lett. Nuovo Cimento} {\bf
4},(1972),7371;  {\it Phys. Rev.}{\bf D7},(1973),2333.}
on the amount of information that can be hidden behind the horizon of a
black hole of mass $M$, is evaded by black holes with horny geometry.
Bekenstein's argument relied on the notion that only particles with
Compton wavelength smaller than the radius of the horizon could ``fit
in'' to the black hole.  But this is untrue for the infinite volume {\it
cornucopion}\horned .  Thus we are no longer faced with the dilemma of
having to explain how all the information contained in a large mass
black hole is emitted in the process of Hawking evaporation down to a
smaller mass hole\foot{It is crucial to this argument that our theory
have modes which propagate as free particles in the two dimensional
horn of the cornucopion.  Information stored in other modes of the field
can be transferred into these waves by interactions near the
throat of the cornucopion.}.

In the present paper, we wish to present more details of the arguments
which led us to believe that cornucopions resolve the coherence loss
problem of black holes.  We begin with a short discussion of the
cornucopion solution in string theory, arguing that the GMGHS solution is
singular in this context.  We then suggest that results on $c=1$ string
theory imply that the GMGHS solution is part of a one parameter family
of solutions, all the other members of which have a nonzero
condensate of massless fermion pair modes.
Since the effective string coupling for scattering within
the horn of the cornucopion is inversely proportional to the strength of
this condensate, quantum corrections to the bare dilaton black hole are
large, but the solutions with large condensate can be described
semiclassically.\foot{Most resummations of the perturbation series in
the $c=1$ model give a finite limit for the S-matrix as the tachyon
condensate goes to zero\ref\moore{G.Moore,R.Plesser, {\it Classical
Scattering in (1+1) Dimensional String Theory},Yale Preprint YCTP-P1-92,
hepth@xxx.lanl.gov - 9203061.}.  Thus, in the quantum theory the
solution is nonsingular, but it is not amenable to semiclassical
investigation.}  We take this dilaton black hole with a large
condensate as a paradigm for the cornucopion.

 In Section 3 we describe the classical formation of near
extremal charged black holes in dilaton gravity coupled to
electromagnetism.  We argue that the cornucopion begins life as a finite
volume dimple on flat space time, whose tip grows without bound.  The
static cornucopion solution is the asymptotic limit of this growing
solution\foot{It is likely that extremal charged matter does not
collapse classically, because of magnetostatic repulsion.  Our
discussion can be taken as a description of near extremal collapse.  The
cornucopion is hypothesized to be the remnant of Hawking evaporation of
this near extremal object.}.  This leads us to the concept of finite volume
cornucopions, which represent the instantaneous
configurations of the near extremal black
hole a finite time after its formation.  These are approximately static
solutions of the equations of motion whose only time dependence is in
the rapidly receding tip.
In
the fourth section of the paper we review various
scenarios for the endpoint of black hole evaporation, concentrating on
the difficulties of scenarios which invoke the existence of relic
particles.

The problems with relic particle scenarios are all a consequence of the
infinite number of approximately degenerate states that must be present
if the relics are to possess the information content of the large black
holes from which they were formed.  In the case of cornucopions this
infinite number of states is associated with the infinite volume of the
horn. We begin section 5 of this paper by expanding on the arguments made
in\horned about the difficulty of bringing cornucopions into thermal
equilibrium with an external heat bath.  We argue that the combination
of these arguments with those given below, which suggest small amplitudes for
cornucopion pair production, remove
the problems posed for thermodynamics by an infinite set of degenerate
states.  We then
study the contribution of finite volume cornucopion
configurations to virtual loops
by heuristic semiclassical techniques, and argue that the
amplitude for this virtual process is of order $e^{-{V\over g_0^2}}$
where V is the cornucopion volume and $g_0$ is related to the value of the
dilaton field at infinity in the usual manner.  We argue that the sum
over the large but finite number of states in the cornucopion corrects
this by a factor $e^{c V}$ where $c$ is a positive constant
independent of $g$.  Thus, if the coupling is small enough, the sum over
cornucopion volumes and internal states converges and cornucopions give only
small finite corrections to low energy scattering processes.  We show
that this argument is consistent with crossing symmetry by drawing an
analogy with the scattering and production of solitons in weakly coupled
field theories.  Order one scattering amplitudes for elementary
particles off solitons are consistent with exponentially small
production cross sections because the production cross section is
related by crossing to a large momentum transfer scattering amplitude.
The particle soliton scattering amplitude can be large for small
momentum transfers but falls exponentially with the momentum transfer
because the soliton form factor is the Fourier transform of a smooth
classical field.  By analogy, we argue that virtual production of pairs of
large
volume cornucopia is related by crossing to processes in which the
cornucopion volume is changed by a large amount due to scattering from
an elementary particle.  These processes are extremely improbable, and
there is no inconsistency with our estimate for the contribution of
cornucopions to virtual loops.

Unfortunately, these semiclassical arguments are not valid for the model
of cornucopions given by the extremal black hole of four dimensional
Einstein-Maxwell-Dilaton gravity.  For this solution, the effective
coupling grows as one proceeds down the horn of the cornucopion, and one
finds a finite production probability by naive semiclassical estimation.
Of course, the semiclassical approximation breaks down in this case and
one must understand strong coupling physics.  We cannot do this at
present for real cornucopions, but some progress can be made by studying
an analogous problem in $1+1$ dimensional string theory where the strong
coupling singularity is shielded by a condensate of massless modes.  In
that model we argue that the classical Euclidean action is indeed of
order the volume.  However, we must then inquire whether the effective
volume of the system is truly infinite.  This turns out to depend on how
one resums the semiclassical perturbation expansion, and at the present
time there do not exist reliable criteria for deciding which resummation
is correct.  Thus, although the case against cornucopia as
remnants of black hole evaporation is unproven, the resolution of the
argument may depend on strong coupling physics.  We will show that if
one takes the pessimistic point of view that strong coupling effects cut
off the horn of the cornucopion, then its residual entropy is of order
its mass, in parametric agreement with the Bekenstein-Hawking formula.

We end section five with a discussion of the real pair production of
cornucopia in weak static magnetic fields.  It has been argued, that in
the weak field limit, the calculations of Affleck and Manton for
solitons\ref\affleck{I.Affleck,N.Manton, {\it Nucl. Phys.}{\bf
B194},(1982),38} and/or Garfinkle and Strominger for Wheeler
wormholes\ref\Garfstrom{Garfinkle,A.Strominger, {\it Phys. Lett.}{\bf
B256}(1991),246.} indicate that cornucopions will be produced at the same rate
as
elementary particles of the same mass in such background fields.  We
give a critical discussion of these arguments, and explain an
alternative semiclassical picture of cornucopion pair production, which
makes the Affleck-Manton estimate of the amplitude for production of the
cornucopion geometry consistent with finite total cross sections.  We
argue that the tunneling process which gives rise to cornucopion pairs,
produces cornucopions with a size of the order of the characteristic
Schwinger length for production of elementary particles of the same mass
as the cornucopion.  These relatively small cornucopions then grow
classically to arbitrarily large size.  In this method of pair
production, the external field cannot create most of the states of the
infinite cornucopion geometry.  It creates only those that can arise
>from initial conditions which ``fit in'' to the original small volume.
As in the inflationary universe, this is a small subset of the total
number of cornucopion states.

 In
the conclusions we note that cornucopions may be only one of a number of
classes of objects that look like small black holes from the outside,
but conceal a large interior world.  We stress that black holes whose
singularity is replaced by an interior DeSitter space (which appear in
the work of Farhi and Guth\ref\Guth{E.Farhi,A.Guth,{\it Phys.
Lett.} {\bf B183},(1987),149},V.P.Frolov,M.A.Markov and
V.F.Mukhanov\ref\Frolov{{\it Phys. Rev.}{\bf D41} (1990),
383},Morgan\ref\morg{D. Morgan, {\it Phys. Rev.} {\bf
D43},(1991),3144.}
and Strominger\ref\strom{A.Strominger,
Santa Barbara preprint UCSB-TH-92-18, hepth@xxx/9205028}) may represent
a particularly attractive endpoint for black hole evaporation.

\newsec{\bf Is the Extremal Dilaton Black Hole Singular?}

Garfinkle, Horowitz and Strominger argued that the extremal dilaton
black hole was nonsingular.  The basis of their argument was that the
apparent singularity was an infinite distance away in the \lq\lq stringy
metric'':

\eqn\metric{ds^2 = - dt^2 + e^{4\phi} d{\bf x}^2}
\eqn\dilaton{e^{2\phi} =  g^2 = e^{2\phi_0} + {2M e^{\phi_0}\over |x|}}

Taking geodesic distance as a predictor of physical evolution is
dangerous in theories which involve scalar fields.  One can always
perform Brans-Dicke transformations in which the metric is Weyl
transformed by a positive function of the scalars.
This is a point transformation on the configuration space of fields, and
cannot change the physics.  However it does change geodesic distances,
in a way that can be singular if the scalar fields develop
singularities.  The physics of the model depends on how all the fields
in the theory are coupled not only to the metric, but also to the
dilaton and other scalars. GHS argued that, since the world sheet
lagrangian of string theory describes geodesic motion in the stringy
metric, this was the appropriate physical measure of distance in string
theory. However, once we begin to calculate scattering amplitudes, the
dilaton field begins to play an important role.  In string theory,
each vertex operator carries a factor of the string coupling constant, which
becomes infinite at the end of the infinite horn of the GHS solution.

There is as yet no known exact conformal field theory representation of
the extremal dilaton black hole solution of heterotic string theory.
However, if we concentrate on physics in the horn of the cornucopion we
can make a plausible guess at some of the features of this conformal
field theory.
The world sheet Lagrangian describing scattering within the horn should
be the world sheet supersymmetric completion of a Lagrangian of the form:
\eqn\twodstring{{\cal L}_{ws} = \partial y {\bar\partial} y - \partial
\tau {\bar\partial} \tau - Q R^{(2)}y + {\cal L}_{compact}}
Here $y= ln x $, $\tau$ is the time coordinate, and
${\cal L}_{compact}$ describes a unitary conformal field theory
with a discrete positive spectrum of conformal dimensions. It represents
the angular degrees of freedom of three dimensional space, as well as
the six compactified dimensions.  We do not know the right and left
central charges of the Lagrangian ${\cal L}_{compact}$, but they
cannot be equal to zero.  The spectrum of the theory then contains
potential tachyons, which (it is to be hoped) are eliminated by the
physical state conditions.

As a consequence of the discrete spectrum of dimensions of ${\cal
L}_{compact}$ most of the
particles in this theory propagate with large effective masses in the horn.
The low energy field theoretic description of the system implies that
there are exceptions to this rule, which
are related to charged fermion zero modes around the
monopole. These states were discussed
in\ref\alford{M.Alford,A.Strominger, {\it S Wave Scattering of Charged
Fermions by a Magnetic Black Hole}, Santa Barbara preprint
NSF-ITP-92-13, hepth@xxx.lanl.gov - 9202075.} and\horned . They
propagate as
massless two dimensional fermions.  The vertex operators for these
states have the form $e^{(\alpha + ik)y + iEt} {\cal O}$, where
 $ {\cal O}$ is constructed from the degrees of
freedom of the ${\cal L}_{compact}$ theory.  On shell vertex operators
satisfy $E^2 = - (\alpha + ik - {Q\over 2})^2 +{Q^2 \over 4} + \Delta
-2$, where $\Delta$ is the dimension of ${\cal O}$.  In order that the
states have real energies we must have $\alpha = {Q\over 2}$, and in
order that they be massless, $\Delta = - {Q^2 \over 4} + 2$.

Inserted into the $y$ path integral, these operators generically cause
divergences in the integral over the zero mode of $y$.  This is
analogous to the divergences of tachyon amplitudes in the $c=1$ matrix
model when the cosmological constant is equal to zero.  In that case,
the cure for the disease is well known.  The zero energy tachyon
vertex operator can be added to the Lagrangian without destroying
conformal invariance.  This tachyon condensate is a new classical
solution of $c=1$ string theory and the perturbation expansion around it
is computable and finite.  Study of that expansion shows that the
effective expansion parameter is $g_{st}^2 \over \mu$ where $\mu$ is the
coefficient of the tachyon condensate in the world sheet Lagrangian (the
two dimensional cosmological constant).  The singularity of the $\mu =
0$ solution is seen as a failure of the semiclassical expansion.
Various nonperturbative resummations of the expansion give a perfectly
sensible S-matrix at $\mu =0$.

It seems rather hopeless to try to find an analogous nonperturbative
solution of the string theory associated with the horn of the
cornucopion.  We do not even have an exact conformal field theory to
start from.
Rather we should look for the analog of the solution with
finite $\mu$.  It does not seem to make sense
to simply put the vertex operators of
the massless modes into the world sheet Lagrangian, for they are
spacetime fermions!  From a spacetime point of view, we would expect a
condensate of these zero modes to be described by the bosonized fermion
current. We have no idea how to represent these scalar fields in terms
of vertex operators in string theory.  Although the bosons are two
fermion states, it is not entirely outlandish to expect to find them in
the one string Hilbert space.  The two to two fermion scattering matrix
should have a pole corresponding to exchange of these bosons.  Since they
are derivatively coupled, it should show up as a $0 \over 0$
contribution to the zero energy scattering amplitudes.  Such structures
in spacetime often have signatures in the behavior of vertex operator
correlation functions near the boundary of moduli space, which can be
associated with other vertex operators\foot{We are thinking of the
Dine-Seiberg vertex operators\ref\ds{M. Dine, I. Ichinose, N. Seiberg,
{\it Nucl. Phys.} {\bf B293}, 253 (1987).} for the auxiliary components of
superfields. These are also composite operators from the spacetime point
of view.}.

We hope to return to these fascinating issues at some future time, for
they are crucial to a semiclassical understanding of the GMGHS solution.
For the purposes of the present paper we will make the optimistic
assumption that an
analog of the nonsingular $\mu \neq 0$ solutions exists,
and that its properties are similar to those of the $c=1$ model.
It is this hypothetical nonsingular solution that we want to use as a
model for the cornucopion. The
careful reader will note that there is only one point below in which we
make use of the nonzero value of the condensate.  Most of the properties
of the cornucopion follow from its geometrical structure, and would be
valid for other sorts of black hole remnants which conceal a large
internal space behind their apparent horizons.

Before closing this section we should draw attention to a possible
problem with the idea of using the $c=1$ string theory as a model for a
cornucopion. S.Shenker\ref\shenker{S.Shenker {\it Private
Communication}} has pointed out to us that if, as one expects in the
cornucopion,  the string coupling goes to a finite constant in the
throat region, the effective volume of the $c=1$ world appears in
perturbation theory to be finite and of
order the logarithm of the string coupling.  In the extremal dilaton
black hole, the volume over which the two dimensional world is weakly
coupled is also
proportional to the mass of the black hole.  Thus if we make the
pessimistic assumption that the strong coupling region is really
inaccessible, the cornucopion will be capable of storing an amount of
information that is bounded by a constant of order its mass.  This bound
is parametrically the same as that given by the Bekenstein-Hawking
formula.  Note however that in many nonperturbative resummations of the
$c=1$ pertubation series, the barrier that prevents tachyons from
penetrating the strong coupling region is finite, and there is another
weakly coupled region on the other side of it which contains an infinite
number of states\foot{The structure is reminiscent of the other world in
the Kruskal extension of the Schwarzchild solution.  The major
difference is that in the c=1 model there is no singularity or horizon
dividing the two worlds, and they can communicate with each other.}.
Thus, the question of whether a nonsingular cornucopion can have an
infinite number of states is bound up with nonperturbative physics.

\newsec{Collapsing Cornucopions}

In this section we describe the classical collapse processes that could
lead to near extremal black holes.
Consider a collapsing shell of magnetically charged matter, in
dilaton-Einstein-Maxwell theory. One can attempt to construct a solution
representing collapse by gluing the
four dimensional exterior solution of GHS onto a smooth interior vacuum
solution of the equations with the topology and symmetry of a 3-hemisphere.
 The spherically symmetric geometry is a two sphere with time and radial
 coordinate dependent radius cross a two dimensional spacetime geometry
 for the $r - t$ submanifold.  We call the radius of the two sphere
 $e^{2 \sigma (r,t)}$ and use synchronous coordinates.  The Lagrangian
 for the most general solution of this form is
\eqn\dilgrav{
S = \int \sqrt{-g} e^{-2\phi} (e^{2\sigma}(- R -
2(\partial\sigma)^2 - 4(\partial\phi)^2 + 8 \partial\phi \partial\sigma)
- 2 + Q^2 e^{-2\sigma})
}
 We have chosen coordinates in which the exterior metric is;
\eqn\extmet{
ds^2 = - dt^2 + {1\over{h(t,r)^2}} dr^2 + e^{2 \sigma(t,r)} d\Omega^2
}

where $h(t,r) = (1 - {Q\over r})$ and $\sigma = log(r)$, for the static
GMGHS solution.

Now suppose that this is the solution outside a
collapsing shell of magnetically charged matter with two-dimensional
world-line $(t,r) = (T(\tau),R(\tau))$, where $\tau$ is the proper
distance along the world-line. To obtain the solution
inside the collapsing shell we use the dilaton gravity action with
zero magnetic field.  The latter condition is a consequence of our
assumption of spherical symmetry.  Inside the shell, there are no
magnetic sources and the field equation for the magnetic field is
\eqn\source{\nabla_{\mu} (e^{-2\phi} F_{\mu\nu}) =0}
Since $\phi$ does not depend on the angular variables, and only the
angular components of the field are nonzero, the field must be constant
inside the shell.  Continuity of the solution at the origin restricts
this constant to be zero.

Introduce co-ordinates using the proper time of the shell, so that
$ds^2 = - d\tau^2 + dn^2$, along the world-line of the collapsing
matter. We will use these co-ordinates to expand the solution for $\phi$
and $\sigma$ in powers of n toward the interior of the shell, with
coefficients of the power series being functions of $\tau$.
We will further simplify the equations by making a change of
variables:
\eqn\change{
u(\tau,n) = e^{\sigma - \phi}
}

In terms of these fields the Lagrangian is,
\eqn\lagus{
S = \int{\sqrt{-g}(-u^2 R + 2 u^2 (\partial \sigma)^2 - 4
(\partial u)^2 - 2 u^2 e^{-2 \sigma} + Q^2 u^2 e^{-4 \sigma})
}}
The equations of motion for this Lagrangian are given in Appendix 1.
Of course, to find interior solutions with zero magnetic field to  match
onto the exterior extremal dilaton
solution, we set $Q = 0$ in the above equation.

The power series  for the fields $\sigma$ and $\phi$,
expanding from the shell towards the interior,
is

\eqn\expandu{
u(\tau,n) =R(\tau)  \sqrt{1 - {Q\over R(\tau)}}(1 + f_1(\tau) n +
f_2(\tau) n^2 + f_3(\tau) n^3 + \dots
}
\eqn\expandsig{
\sigma(\tau,n) = log(R(\tau)) + d_1(\tau) n + d_2(\tau) n^2 +
d_3(\tau) n^3 + \dots )
}

where the coefficients of the leading terms are determined by
continuity of $\phi$ and $\sigma$ across the shell. The metric is
$g_{\mu\nu} = diag(-h(\tau,n)^2, g(\tau,n)^2)$, and the coefficients
have the expansion;

\eqn\expandh{
h(\tau,n) = 1 + h_1(\tau) n + h_2(\tau) n^2 + h3(\tau) n^3 + ...
}
\eqn\expandg{
g(\tau,n) = 1 + g_1(\tau) n + g_2(\tau) n^2 + g_3(\tau) n^3 + ...
}

The equations that we have to describe this system now consist of the
equations for $u$ and $\sigma$, and the stress tensor equation. At the
boundary of the collapsing shell there is a non-trivial matching
equation for the stress tensor component $T_{00}$ \foot{see Appendix
2 for details}.

We will assume that the classical Lagrangian for the matter that
constitutes the shell is of the form,
\eqn\lagmatt{
S = \int{\sqrt{-g} u^2 (-(\partial A)^2 - m^2 A^2 + \dots)}
}
That is, the matter in the collapsing shell couples to the dilaton like
some massive mode of the string.
In the rest frame of the collapsing
shell, the matching equation reads\foot{The full details of the
derivation
are in Appendix 2.};
\eqn\matcha{
M u(\tau,0)^2 = \int_{-\epsilon}^{\epsilon}{T_{00} dn}
}
which becomes;
\eqn\matchb{
M u(\tau,0)^2 = R (1 - {Q \over 2R})\sqrt{\dot{R}^2 + (1 - {Q \over R})^2}
- u \partial_n u(\tau,0)
}

At this point we must be more specific about the fields on the interior
of the shell.  In Einstein's theory, there is a unique spherically
symmetric nonsingular vacuum solution, but here the dilaton dynamics
gives rise to an infinite set of spherically symmetric solutions of the
source free field equations in a finite region.  We have tried to
restrict the solution by assuming a cosmological form for the metric,
$ds^2 = -d\tau^2 + a(\tau)^2(dr^2 + r^2 d\Omega^2)$, inside the shell,
but this is inconsistent with the field equations.  Similarly, an
attempt to keep the three dimensionally conformally flat form of the
metric, with conformal factor tied to the dilaton, is inconsistent.  We
have not been able to come up with a natural ansatz.  Nonetheless, we
believe that smooth solutions exist.  There are many smooth solutions of
the vacuum field equations restricted to a manifold with the topology of
a hemi-3-sphere cross time.  Our matching conditions fix only the values
of the metric functions and dilaton along the timelike world line of the
collapsing shell, leaving their normal derivatives undetermined.  Thus
there seems to be plenty of room for patching in a nonsingular vacuum
solution.

To obtain some feeling for the motion of the collapsing shell we
have made the fairly arbitrary assumption that:
\eqn\assa{
f_{1}(\tau) = {a \over R(\tau)}
}
This gives us a
single first order ordinary differential equation
for $R(\tau)$\foot{Appendix 3}. The solution
so obtained behaves like $R(\tau) \approx Q + e^{- \gamma \tau}$, as
$\tau \rightarrow \infty$. We can then use this solution to check that
the behaviour of the other coefficient functions,
to leading order, are well behaved for all finite values of
$\tau$. We can continue this procedure perturbatively, to verify that
the coefficients in the expansion in powers of $ n$ are smooth
functions of $\tau$.  Of course, this demonstration of a smooth
perturbation expansion around the shell, does not guarantee the
existence of an everywhere smooth solution.
We continue to search for a sensible ansatz that will enable us
to demonstrate explicitly the existence of a smooth collapsing solution,
but we feel confident that such a solution exists.

The collapsing solution that we have described, begins as a dimple on
flat space.
At any finite time after its formation, it
will have the geometry shown in Figure 2.  We will refer to such an
object as a finite volume cornucopion.  It is a solution of the field
equations that is static over most of space.  The time dependence occurs
only in the tip of the horn.

\newsec{\bf The Problem of Stable Relics $\ldots$}

Since the publication of Hawking's seminal papers on black hole
evaporation 18 years ago\ref\hawking{S.Hawking,\ {\it Nature}\ {\bf
248},(1974), 30;\ {\it Phys.Rev.}{\bf D14},(1976),2460.}, there have
been many attempts to
resolve the puzzle of information loss that is apparently implied by the
Hawking process.  In broad terms, these attempts fall into three
classes.\foot{S.Giddings\ref\giddings{S.Giddings,\ {\it Black Holes and Massive
Remnants}, Santa Barbara Preprint UCSBTH-92-09, hepth@xxx/9203059.}
 has recently presented a
 discussion of scenarios
for the endpoint of black hole evaporation which overlaps
 with this section.}  In the first, essentially that originally advocated by
Hawking, one accepts the information loss at face value and attempts to
describe processes in which pure states can transform into mixed states.
This idea raises a number of paradoxes\ref\bps{T.Banks, M.E.Peskin,
L.Susskind, {\it Nucl. Phys.}{\bf B244} (1984),125.},
and it is not clear\foot{At
least to the present authors} that it can lead to a sensible physical
theory.

Advocates of the second approach, including Page, 't Hooft, Wilczek, and
Susskind and their collaborators\ref\simple{D. Page, {\it Phys. Rev.
Lett.} {\bf 44},(1980),301;
G. 't Hooft, {\it Scattering
Matrix for a Quantized Black Hole}, p. 381-402, {\bf Black Hole
Physics}, V.De Sabbata and Z.Zhang (eds.), Kluwer Academic Publishers,
the Netherlands, 1992; L.Susskind,L.Thorlacius,\ {\it
Hawking Radiation and Back-Reaction}, Stanford preprint SU-ITP-92-12,
hepth@xxx.lanl.gov-9203054; \ A.Peet,L.Susskind,L.Thorlacius,\ {\it
Information Loss and Anomalous Scattering}, SU-ITP-92-16;\
G.'tHooft,\ {\it
Nucl. Phys.}\ {\bf B335},\hfil\break (1990),138, and references therein;\
F.Wilczek,C.F.E.Holzhey,\ {\it Black Holes As Elementary
Particles},Institute for Advanced Study preprint, IASSNS-HEP-91-71.}
insist that the
information loss is a consequence of an improper treatment of the
quantum mechanics of the gravitational field.  They argue that in a more
careful analysis, which goes beyond the semiclassical approximation,
Hawking radiation will be shown to carry information,
encoded in subtle nonlocal correlations, much like those in the \lq\lq
thermal'' radiation emitted from any hot body.  These authors face the
challenge of understanding the entropy produced
in Hawking's calculation of the decay of a large classical black hole
into one of half the mass.  According to the Bekenstein-Hawking formula
for the entropy of a black hole of given mass, the Hawking radiation
emitted in this process carries a huge entropy $\sim {M^2 \over M_P^2}$,
but according to the second point of view it actually carries none.
Thus there must be a large corrections to Hawking's calculation of the
density matrix of the emitted radiation, despite the fact that the
entire process takes place within what appears to be the domain of
validity of the semiclassical approximation.  If one takes the
semiclassical calculation seriously in regions inside the horizon but
away from points of high curvature, one is led to serious problems of
causality. The information carried by the infalling matter is still
localized behind the horizon on a spacelike surface on which most of the
mass of the black hole has been radiated away.  In addition to this, the
subtle correlation approach can never account for the global quantum
numbers that appear to be lost in black hole decay.  One is led to claim
that black hole physics can only make sense in the context of a theory
in which there are no conserved global quantum numbers.

The final approach to the problem of Hawking radiation is to postulate
the existence of an infinite number of stable remnant objects, all of
whose masses are sufficiently small that the Hawking calculation breaks
down near the corresponding Schwarzchild radii.  These objects can store
any global quantum numbers that have fallen down the black hole, and the
infinity of degenerate states is a repository for the information lost
in the Hawking process.  A recent critical discussion of this scenario
can be found in\ref\casher{Y.Aharonov,A.Casher,S.Nussinov, {\it Phys.
Lett.} {\bf B191}, (1987), 51.}.

The problems of the stable relic scenario are all caused by the infinity
of degenerate relic states that it requires.  All formulas of
statistical mechanics are formally
infinite (even in the microcanonical ensemble)
if we assume that these states have come into equilibrium with the rest
of the world.  The same can be said for all formulas in quantum field
theory, when loops of virtual relic particles are taken into account (at
least if we treat the relics as elementary particles).
Furthermore, if the probability of pair production of a single relic
state in an external environment (say a weak slowly varying
electromagnetic or gravitational field) is bounded from below by a
positive number $\epsilon$, no matter how small, then the total
production probability is infinite.  Typical pair production cross
sections for elementary magnetically charged
particles, or monopole solitons, in static
magnetic fields go like $e^{- {M^2 \over g B}}$ where
$M$ is the mass of the state being produced.  If the infinite set of
black hole relics can all be produced at rates of this order of
magnitude, then everything in the world will decay into black hole
relics in a microscopic time.  Given these striking conclusions, it is
not surprising that many authors prefer the scenario in which
information is emitted with the Hawking radiation to the relic particle
scenario.

\newsec{\bf $\ldots$ And Its Resolution?}

We have just outlined the obstacles faced by any attempt to
identify stable relics as the endpoint of Hawking evaporation.  In this
section we would like to show how cornucopia overcome these obstacles.
A brief discussion of the thermodynamic equilibration of cornucopia has
already appeared in \horned , so let us begin by expanding on it.
Imagine that we have created a single cornucopion in one of its many
degenerate quantum states, and that we have filled the universe outside
it with a gas of particles at temperature $T$.  Let us ask how long it
takes for some subset of states of the cornucopion to come into thermal
equilibrium with the external gas.  To begin with, let us imagine that
interactions of the gas near the mouth of the horn can excite states
inside the cornucopion with probability $\sim e^{- {E_{ADM}\over kT}}$
where $E_{ADM}$ is the Arnowitt Deser Misner energy measured by an
observer in the asymptotically flat region of spacetime.  From causality alone
we have a restriction that in time $t$ we can at most excite states of
the horn within a distance $t$ of its mouth.  The number of such states
is of order $e^{t M_P}$ and their contribution to the partition function
is at least $e^{t M_P - {M_P \over T}}$ if we assume that typical
splitting in ADM energy between these states is of order the Planck
mass\foot{If most of the states have zero ADM energy, the temperature
dependent factor is absent.}.
Since the time $t$ over which one can imagine thermalization to take
place in an expanding universe is always bounded from above by the
inverse of the expansion rate, $H^{-1}$ we never have to think about a strictly
infinite number of states.

In fact it is highly unlikely that all of the states within a distance
$M_P \over H$ of the cornucopion throat will be thermally distributed.
There are two basic reasons for this.  Firstly, there are large
repulsive potential barriers located near the cornucopion throat for
most modes of the fields in the external universe\foot{This is shown for
example in the paper of Holzhey and Wilczek cited in\simple , and was
also apparently known to GHS.}.  In
string theory this can be understood from the statement that the
effective world sheet theory which describes the horn
of the cornucopion is the tensor product of the the Liouville theory and a
compact conformal field theory.  Thus, apart from the small number of
massless
modes described above, all states have effective masses
in the cornucopion horn that are of order the string scale.
The probability for an external particle to penetrate the throat
and produce an ingoing wave of these massless modes
is very small; at least proportional
to inverse powers of the Planck scale.\foot{One should avoid being
confused by the strong Callan-Rubakov interaction of external particles
with a magnetically charged cornucopion.  This describes processes going
on very far (in Planck units) from the cornucopion throat.}
Thus the thermalization of the cornucopion proceeds at a very slow rate,
likely to be slower than the expansion rate of the universe at most
epochs of interest.

Our second reason for believing that cornucopion states
will not come into conventional thermal equilibrium is more difficult to
explain because we do not fully understand it.  Imagine that some
process in the external world has succeeded in injecting a pulse of
tachyons into the horn of the cornucopion.
Our analogy between the horn of the cornucopion and the
$c=1$ model tells us that (at
least to all orders in the string loop expansion) the dynamics of this
pulse is describable by a two dimensional field theory of interacting
massless particles.  The question of which states of this field theory
are occupied seems to have little to do with the ADM energy measured by
an observer in the asymptotic four dimensional region of space time.
Its dynamics are governed by an effective Hamiltonian ${\cal H}$
whose connection with
ADM energy is far from clear.  One might expect that after enough time,
the state of the system interior to the horn would be well described by
a density matrix of the form $e^{-{{\cal H} \over T_h}}$ where $T_h$ is an
internal temperature whose relation to the temperature of the external
universe is less than obvious.  However, if a finite pulse of massless
particles
({\it i.e.} the analog of a
 distribution of tachyons that can be described by a finite perturbation
of the fermi surface in Polchinski's\ref\jopo{J.Polchinski,{\it Nucl.
Phys.}{\bf B362},(1991),125.}
 description of the tachyon field theory of the $c=1$ model)
was injected, this temperature will be finite.  This is not a thermal
distribution in ADM energy.

We do not pretend to fully understand these arguments (particularly the
latter) so it is fortunate that even if the states of the cornucopion
had come into thermal equilibrium they would have little effect on the
thermodynamics of the external world.
The effect of the extra states of the cornucopion is to endow {\it
each} of the states of the world external to the black hole with an
enormous degeneracy.  Let us use the letter $\alpha$ to label this
degeneracy.  Now consider any operator ${\cal O}$, localized in a region
R of space time external to the cornucopion.  We claim that ${\cal O}$
is essentially the unit operator as far as the $\alpha$ label is
concerned. This argument can be stated more precisely.  In \horned we
argued that the degeneracy in ADM energy of states of the cornucopion
was a consequence of the existence of states concentrated in regions far
down the horn.  The energy difference between two states labelled
$\alpha$ and $\alpha^{'}$ goes to zero exponentially as the
difference between the two states is taken to be a state localized
further and further down the horn.  But in the same limit, the
difference between the expectation values of ${\cal O}$ in these two
states vanishes, as does its off diagonal matrix element between them.  Thus
\eqn\thermexp{\sum_{N,\alpha}<N,\alpha |{\cal O}|\alpha ,N> =
\sum_{\alpha} \sum_{N}<N |{\cal O}|N>}
The sum over $\alpha$ factors out when we compute expectation values.
Thus for local quantities located far enough from the cornucopion,
thermal averages do not probe the presence of its large number of
degenerate states.  We emphasize that there is nothing exotic about this
argument, and that it depends principally on the fact that the interior
of the cornucopion is {\it far away} from the external observer.  If
someone increases the local density of states on the moon, it effects
thermodynamics on the moon, but not on earth.

The effect of cornucopia on thermodynamics in regions far from the
throat of the black hole is thus seen to be rather innocuous. We believe
that the widespread belief that infinite numbers of black hole
remnants,degenerate in ADM energy,
contradict thermodynamics is based on the misconception that these
remnants could be described as particles, and that one should be able
to produce them all in the laboratory. The fundamental problem of
thermodynamics in the stable relic scenario is thus intimately related
to the estimate that we made in the previous section of production of a
stable relic, thought of as an elementary particle.

We are thus led to investigate the fundamental problem of any relic
scenario, the pair production of cornucopia, and their
contribution to virtual loops.  The most serious sounding argument
against the existence of stable relics of black holes is based on
crossing symmetry.  The scattering of low energy photons and
gravitons from a cornucopion is completely determined by the object's mass and
charge, and the scattering amplitudes are not small.  Crossing symmetry,
it is argued, should related these amplitudes to cornucopion production
amplitudes, and (by unitarity) to their contributions to virtual loops.
If each state of the cornucopion is produced with finite amplitude,
and/or gives a finite contribution in loops, the sum over the infinite
number of virtual relic states will give rise to infinite production
cross sections and infinite renormalizations of all low energy
amplitudes.

We will argue in a moment that this argument does not take sufficient
account of the infinitely extended nature of the cornucopion geometry.
However, it is worth pointing out that such arguments from crossing
symmetry can be misleading even for extended objects of finite extent,
namely solitons in ordinary weakly coupled local field theories.
In particular, naive application of crossing symmetry would lead us from
the finite Thompson cross section for low energy photon scattering off
an 't Hooft Polyakov monopole, to the conclusion that the production
cross section for such monopoles above threshhold in electron positron
annihilation was of order one.  From unitarity we would then conclude
that monopoles contributed to photon vacuum polarization at some finite
order in perturbation theory (some inverse power of the monopole mass).
In fact, as argued long ago by Drukier and
Nussinov\ref\nuss{A.Drukier,S.Nussinov,{\it Phys. Rev. Lett.}{\bf
49},(1982),102.} the
monopole production cross section is of order $e^{-{c\over\alpha}}$.
The essence of their argument was that in a weakly coupled theory, it
costs a power of $\alpha$ in probability to create a quantum of the bare
fields from the vacuum.  The coherent monopole state contains of order
$1\over \alpha$ quanta.

To understand where the crossing symmetry argument went wrong, we note
that the Thompson cross section is related to the form factor of the
electromagnetic current (we suppress irrelevant Lorentz indices and
kinematic factors)
\eqn\formfac{<p+q|J|p> = F (q^2)}
at small spacelike momentum transfers $q^2$.  The monopole production
cross section on the other hand is, by crossing symmetry, related to the
analytic continuation of the same function $F$ to large timelike $q^2$
of order the monopole mass $\sim {m_W \over\alpha}$.  A better crossing
symmetry estimate of the production cross section would be to relate it
to the behavior of the monopole form factor at large spacelike $q^2$
where it falls exponentially (it is essentially
the Fourier transform of the smooth monopole field
configuration).\foot{Even this estimate is suspect, though it gives a
more plausible answer.  One could encounter Stokes lines in the analytic
continuation from large spacelike to large timelike $q^2$. However there
is an intuitive connection betwen this estimate and the Drukier Nussinov
argument. The smoothness of the high energy monopole configuration is a
consequence of the fact that it contains a large number of soft quanta.}

We have introduced this example not only because it illustrates how a
naive application of crossing symmetry can grossly overestimate the
production rate of an extended semiclassical object, but because we
believe that it provides a good analogy to cornucopion production, with
the cornucopion volume playing the role that momentum plays in the
monopole form factor.  The conventional pictorial argument for crossing
symmetry is shown in Figure 3.  In the left half of this figure, a
particle is produced by some classical apparatus, scatters off a photon,
and is absorbed by another classical machine.  The amplitude is nonzero
even when the emission, absorption and scattering events are in
spacelike relation.  Thus, some Lorentz observers can see the scattering
occur before the emission or absorption, and interpret the amplitude as
a production process.

In \horned we associated a particlelike coordinate variable
$x^{\mu}(\tau )$ to a single cornucopion.  The dynamics of the system
involved a conventional functional integral over this variable, and so
it is still presumably true that the photon cornucopion scattering
amplitude is nonvanishing when the events $A$, $B$ and $C$ are in a
spacelike relation.  However, the meaning of this picture is quite
different for cornucopia and for particles.  Event $A$ is a classical
cornucopion production event.  We can imagine creating it by letting
some matter collapse into a black hole, possibly followed by Hawking
radiation to produce a cornucopion with Planck sized throat.  It is thus
a violent classical process, not a microscopic event.  Event $B$
is an interaction of the photon with the throat of the cornucopion, it
is presumably much more complicated than the corresponding interaction
with a point particle, but involves no new issues of principle.
However, event $C$ has a drastically different interpretation for
cornucopia and elementary particles.  Event $C$ is pictured as occuring
at a point in spacetime.  However, it symbolizes the destruction of the
cornucopion, which involves the shrinking of its horn back to flat
space.  The tip of the horn is however a finite and presumably very
large spacelike distance away from event $B$.  Given the initial
conditions in which the cornucopion was formed by gravitational
collapse, we find it unlikely that its classical motion will ever shrink
the horn back to zero size to agree with the configuration assumed in
event $C$. Even if this could happen, it must take a long time.
Thus, in order for the
cornucopion to classically disappear at event $C$,
$C$ must be far in the future of $B$, and their is no confusion about
the time ordering.  The alternative is, that between points $B$ and $C$,
a quantum tunneling event occurred in which the volume of the
cornucopion shrank to zero.

Thus crossing symmetry relates photon pair production of cornucopia not
to simple scattering processes in which the cornucopion's horn is
unaffected, but to processes in which the volume of the horn changes
drastically.  In the analogy to monopole production discussed above, we
would say that ordinary scattering measures the cornucopion ``form
factor at small volume change'', while the production process is
obviously related to scattering with large volume change.
To estimate the probability of such large changes in the geometry we
will first attempt to use semiclassical ideas.

We will first discuss the contribution of cornucopia to virtual loops.
The positive energy theorem assures us of the semiclassical stability of
flat space in quantum gravity.  There should be no exact instantons
describing the decay of flat space into another geometry.What we must do
is to try to estimate the probability for a virtual process, rather than
a tunneling process which terminates in motion through an allowed
classical region.  Thus we want to calculate the Euclidean action for a
virtual transition from flat space to a specified configuration of the
geometry, which then subsides into flat space.  We will take the
intermediate configuration to be the finite volume cornucopion discussed
in section 3, and treat the volume of the horn as a collective coordinate.
  The full amplitude for this virtual process will then take the form:
\eqn\tunnamp{A \sim \int dV e^{- {S(V)\over g_0^2}} Z(V)}
where $S(V)$ is the Euclidean action for the process, $g_0^2 = e^{- 2
D_0}$ is the coupling constant written in terms of the value of the
dilaton in the asymptotically flat region of spacetime, and $Z(V)$ is the
partition function for small fluctuations around the instanton.  To all
orders in the loop expansion, $Z(V)$ will behave as $e^{N M^3 V}$, for
large $V$, where $M$ is the cutoff scale (the string tension scale in
string theory) and $N$ is a number that goes to a constant as $g_0
\rightarrow 0$.  We are assuming that the theory is weakly coupled at
all length scales, and in particular that (as in string theory) the
cutoff is smaller than the Planck Mass by a factor of $g_0$.
In the weak coupling limit, $Z(V)$ counts the number
of states of the cornucopion of volume $V$.  Its logarithm is extensive in the
volume.

In this way of organizing the calculation we see how the infinite number
of states of the cornucopion can sum up to give a finite result in
virtual loops.  The static cornucopion is an idealization.  Its infinite
number of states come from its infinite volume.  Any cornucopion that
was created a finite time in the past will only have expanded to a
finite but enormously large volume.  It will have a finite number of
states that grows exponentially with its volume.  In virtual loops we
sum over these finite volume cornucopions.  An infinite contribution
will be avoided if the integral over $V$ converges.  This will occur if
the coupling $g_0$ is small enough, and if the cost in Euclidean action
to create a cornucopion of volume $V$ grows at least as fast as $V$ for
large volumes.  In an ordinary quantum field theory, it would be
essentially obvious that the action for creating a configuration which
differs from the vacuum over a volume $V$ will be of order $V$.  In string
theory, or the low energy dilaton gravity theory which it gives rise to,
the dilaton field whose exponential multiplies the classical action
density, varies linearly along the cornucopion.  Regions far from the
cornucopion throat give exponentially small contributions to the action
of the static solution.  Thus, for the extremal black hole solution of
GHS it is not clear that
the action for the virtual cornucopion creation process
is proportional to $V$ for large volume.
However, we have argued above that the GHS black hole
is not a good model for the generic
cornucopion solution of string theory. In that context it is singular,
and we argued that the nonsingular solutions have a condensate which
prevents the massless modes in the horn from reaching the strong
coupling region.  In order for this to work, the condensate vertex
operator (which is the field that appears in the spacetime action) must
grow in the strong coupling region.  For example,
in the low
energy effective action for two dimensional string theory,
the tachyon appears as
\eqn\ltac{{\cal L}_T = \ha e^{-2D}\sqrt{|g|} [- (\nabla T)^2 + T^2 ]}
The tachyon condensate increases precisely like $e^D$ as we proceed down
the horn of the cornucopion, so its contribution to the action is
proportional to the volume of the horn.  Although we have no reason to
trust the detailed predictions of the low energy action, they should be
valid in the bulk of the cornucopion where tachyon interactions are
unimportant.  Thus in the $c=1$ model, the spacetime action of the condensate
fields is proportional to the volume of the cornucopion.
If we assume that the same is true for the hypothetical nonsingular
cornucopion solutions described above, we may
 conclude that for sufficiently small coupling, flat
space is stable against decay into cornucopia, and their contribution
to virtual loops is finite.

Note that in the above argument we have assumed that the Euclidean
action
density for the process of cornucopion production is positive.  This is
of course untrue in the naive definition of Euclidean quantum gravity.
It is our belief that a correct treatment of tunneling in quantum
gravity requires one to analytically continue the fields in the
Euclidean action, perhaps in the manner advocated by Gibbons Hawking and
Perry, in such a way that the action is positive and tunneling
amplitudes always correspond to suppression. Such a continuation would
also explain the apparent discrepancy between the fact that the infinite
cornucopion has finite mass and our contention that Euclidean processes
which create it as an intermediate state have infinite action.  The
energy density in a theory of gravity is not positive and bulk
contributions to the energy cancel.

We would like to emphasize that the result of the foregoing calculation
is of great interest in string theory even if our contention that
cornucopions resolve the Hawking puzzle is incorrect.  There is little
doubt that string theory has classical solutions that are highly charged
magnetic black holes with the geometry of a cornucopion.  The above
calculation can be viewed as a weak coupling estimate of the probability
for spontaneous production of a pair of such objects from the vacuum.
Our result indicates that flat space is stable against decay into such
objects for sufficiently weak coupling. Its failure to show such
stability would at least mean the breakdown of the semiclassical
argument for the stability of flat space, and might indicate a true
instability.  Indeed, the calculations suggests a potential
instability at larger values of the coupling, but this cannot be
verified without more detailed knowledge of the function $N(g_0 )$.
Known results in the $c=1$ model suggest that there may be a constant
contribution to $N$ and that it will not vanish as $g_0$ gets large.
The only degrees of freedom that are massless in the horn of the
cornucopion are free throughout most of its volume and give a coupling
constant independent contribution to the free energy density.  It seems
unlikely that this will be precisely cancelled by massive degrees of
freedom, so a strong coupling instability of the flat vacuum seems like
a definite possibility.
Perhaps the dynamics of virtual cornucopion production contributes to the
mechanism by which the string coupling constant is fixed.

We now come to what is probably the most vexing problem for the
cornucopion scenario, real cornucopion production in external magnetic
fields.  The conventional wisdom on this subject goes back to a
paper\affleck by Affleck and Manton on the pair production of 't
Hooft-Polyakov monopoles.  Their argument may be caricatured as follows.
The pair production of point magnetic monopoles in a constant magnetic
field is described by an instanton in the one particle quantum mechanics
which describes single particle motion.  If the field points in the
$x_3$ direction, the instanton is a circle in the $x_3 - x_4$ plane in
euclidean space.  The radius of the circle is determined by the
tunneling condition $R B = 2 m$, where $m$ is the particle mass and $B$
the strength of the magnetic field.

Affleck and Manton show that there is a field theory instanton which
is better and better approximated by this quantum mechanical instanton
in the limit that the magnetic radius $R$ is larger than all length
scales describing the structure of the monopole soliton.  The basic idea
is that if $\phi_0 (x_1 , x_2, x_3)$ represents the static soliton
solution, then it is also a solution of the Euclidean equations of
motion representing a soliton that exists for some length of Euclidean
time.  Now consider $\phi (x) = \phi_0 (x_1 , x_2, \sqrt{x_3^2 + x_4^2}
- R) $ which respresents a soliton moving around an Euclidean circle of
radius R.  This configuration is not an exact solution of the Euclidean
equations, but the leading correction to the equations of motion is
cancelled in the presence of the background field if $R$ is chosen to
take on the Schwinger value $2m/B$ (where $m$ is given by the soliton mass).
Thus $\phi$ can be chosen to be the first term in an expansion of an
exact solution in powers of $L/R$, where $L$ is a characteristic
dimension of the soliton.  Affleck and Manton show that in the limit
$L/R \rightarrow 0$, the action for the solution is given by the usual
Schwinger formula for point particles, plus a correction coming from the
Coulomb interaction which is negligible for weak coupling.

While it is tempting to conclude that the same procedure is immediately
applicable to cornucopions, and that they are consequently produced
(when we sum over states) at infinite rates, there are several
objections to such a conclusion.  The first, as we have emphasized, is
that purely semiclassical considerations, applied to the GMGHS soliton,
are invalid, and the L dependence of the amplitude that we have
predicted is only expected to arise semiclassically
 for a truly nonsingular classical configuration.

Secondly, the Affleck Manton calculation gives us the leading behavior
in an expansion of the instanton action in powers of the
external field multiplied by length scales associated with the structure
of the object.  In order $B^{-1}$ the structure dependence is all
encoded in the soliton mass, and the soliton behaves like an elementary
particle.  In general we might expect a structure dependent term of
order $B^0$.  Structure dependence would appear as a dependence of this
coefficient on the
ratio of scalar and gauge couplings which was not a simple function of
the monopole mass.
This term is absent for 't Hooft Polyakov monopoles
because of the symmetry of the instanton configuration.\foot{There is a term of
order $B^0$ in the monopole production rate, but it comes from Coulomb
interactions and is structure independent.}  The
Garfinkle-Strominger\Garfstrom calculation of Wheeler wormhole
production reveals a term of this order proportional to the Euler
character.  If other curvature squared terms were added to the action,
modifying the structure of the instanton without modifying the behavior
expected for point monopoles, we would obtain other
corrections to the action of the same order. These corrections
 could be interpreted
as dependence on the monopole structure.
It seems to us that the
term in the action
proportional to cornucopion length which we have suggested would be
present, will show up in the coefficent of this order zero term in the
Affleck-Manton expansion, if it shows up at all.   The Affleck-Manton
expansion assumes that all length scales characterizing the soliton are
smaller than the magnetic length $2m/B$, which is surely not valid for
infinitely long cornucopions.   Thus, the question of whether the action
has a term proportional to $L$ can be reliably studied in this expansion
only for $L$ much smaller than the magnetic length. As we study
the production of
cornucopions of various sizes, the extensive term should first appear as
$L$ dependence of the coefficent of $B^0$.   Then, when this term begins
to dominate, the expansion breaks down. Thus we believe that the
Affleck-Manton approximation is invalid for estimating the probability
of tunneling to a virtual geometry of size larger than the magnetic
length.  However, we can imagine a process of tunneling to a cornucopion
of a size within the domain of validity of the Affleck-Manton calculation,
which then expands classically to become infinitely long.  We will
discuss this process in a moment.

Our final objection to the conclusion that the Affleck-Manton argument
is applicable to cornucopia is much less clearly formulated.  The
Affleck-Manton instanton is topologically trivial (relative to the
constant background field), while its analog for cornucopions is not.
The corresponding Euclidean space time has an extra boundary.  Thus
there is no clean argument that such instantons should be included in
the path integral\foot{And of course, no clean argument that they should
not.  Note that while the Garfinkle-Strominger instanton is also
topologically nontrivial, it can be made to look more and more like a
Melvin universe by increasing the external field.  The topologically
nontrivial part of the manifold (the maximal size of the wormhole neck),
becomes metrically smaller and smaller.  With a
little bit of coarse graining we can make it disappear.  Thus there is a
sense in which this configuration is nearly continuously connected to
the Melvin background.  There is no corresponding coarse grained sense
in which the circular deformation of the static cornucopion is
continuously connected to the vacuum.}.
By contrast, the sort of instanton depicted in figure 5, is a
smooth deformation of flat Euclidean space.  We have depicted this
instanton as a sequence of time slices in which virtual cornucopion
production proceeds by dimple formation and growth much as real
formation by gravitational collapse does.  Note that, in keeping with
our prejudices about the dependence of Euclidean action on cornucopion
length, we have imagined that the cornucopion configuration produced by
the external field will have a finite length $L$ at the point when the
virtual state ``pops into existence''.  We believe that $L$ will be of
order the magnetic length $2m/B$.

The cornucopions produced by the above process have finite volume but
can now evolve classically into infinite volume objects.  Classical
evolution involves no further suppression of the amplitude, so we must
inquire whether we have demonstrated a method of generating an infinite
number of states with a fixed amplitude per state.  We believe that we
have not.  The initial classical configuration of the cornucopion
produced by an instanton like that shown in figure 4. is of finite
volume.  Even if we assume equal amplitudes for producing all of the
excited states in this volume, the number of available initial states is
finite and of order $e^L$\foot{Here, as always, we assume an ultraviolet
cutoff on the number of states in finite volume}.  This is much smaller than
the essentially infinite number of states of the final cornucopion.
Thus, we agree with the claim that cornucopia will be
pair produced in an external
field with about the amplitude suggested by the Affleck-Manton
calculation.  However, the method of production will be to tunnel to a
cornucopion whose length is of order the magnetic length, and then
inflate classically. The set of initial data for this classical evolution
is too small to populate the large and ever
growing set
of states of the final geometry.

Our argument here is reminiscent of the argument that inflation solves the
homogeneity problem of the Big Bang.  In an inflationary universe, an
initial microsopic domain becomes much bigger than the currently
observable universe.  Any state of the initial universe becomes
essentially homogeneous in the final geometry.  Thus the full space of
possible initial states of the domain are in one to one correspondence
with the very small subspace of the possible states of the inflated
universe, namely those which are homogeneous.  This sort of expansion of the
Hilbert space would seem to be an essential ingredient of a quantum
theory of geometry.

We make no pretense that the above arguments are conclusive or rigorous.
We believe however that the knee jerk reaction that cornucopions behave
in every way like particles because they look like particles to an
external observer are on no more solid a footing.  This is a true
statement for solitons in an ordinary Lorentz invariant quantum field
theory, but we believe that it is much less than obviously true in a
theory in which geometry is dynamical.  Unfortunately, the GMGHS
solution does not lend itself to semiclassical analysis that would
enable us to settle this question by a simple computation.

\newsec{Conclusions}

There are a number of pressing issues which must be addressed
in order to satisfy ourselves that cornucopions are really a
solution of the Hawking problem.
Firstly, it is imperative to find a way establish the existence
of the nonsingular solution with condensate upon which our
considerations were based.  This is not so much a problem of finding an
exact conformal field theory for the GMGHS solution, as one of
understanding spacetime bosonization in a string theoretic language.
Are there vertex operators which represent the bosonic particle states
formed by two parallel massless fermions?  Is there a classical solution
of string theory which corresponds to a nontrivial static background of
these bose fields, and are its physical properties similar to those of
the $c=1$ model?  Or is the formation of the
fermion condensate not describable in classical language?  Classical or
not, what are the properties of the condensate?  Most particularly, does
the resulting state of the full theory have infinite volume like the
classical solution, or finite effective volume like the linear dilaton
electrodynamics of\ref\peet{A.Peet, L. Susskind, L. Thorlacius, Stanford
preprint SU-ITP-92-16}.  The latter scenario would solve the problem of
infinite numbers of degenerate states in string theory, but would
eliminate cornucopia as candidates for black hole remnants.  The
resulting object would have finite volume, of order its Schwarzchild
radius and information content consistent with the Bekenstein-Hawking
bound.
Cornucopia will resolve the black hole information paradox only if, like
the ``symmetric'' solution of the $c=1$ matrix model, they are truly
infinite spaces.

We should note that the extremal dilaton black hole is not the only
example of a possible remnant object with infinite volume hidden behind
an apparently particulate facade.  In their discussion of creation of
a universe in the laboratory, Farhi and Guth\Guth noted that a
small patch of inflating universe looks to an outside observer like a
black hole.  If the neck that connects the inflating bubble to external
flat space remains of finite extent,
this would be a remnant object with an infinite number of
states.\foot{The idea that such objects could be the endpoint of black
hole evaporation was first presented to us by Michael Douglas.}  Strominger's
``decoupled ghost'' model\strom, of semiclassical two dimensional
gravity seems to contain remnants of this type.  Perhaps if one could
find a four dimensional model with remnants of this type, reliable
semiclassical computations of production amplitudes
could be performed.  It seems plausible that
the relevant instanton would be a piece of
Euclidean four sphere of finite radius (determined perhaps by the total
ADM mass) attached to flat space.  This would correspond to the
nucleation of a finite size universe which would then undergo classical
inflation. As in our discussion of the previous section the nucleation
process could create only a small number of the states of the final De
Sitter universe.  However, if the singularity of a black hole were
replaced with such an internal De Sitter universe, it could serve as a
repository for the information that is apparently lost in the Hawking process.
This is perhaps a more general and attractive scenario for the end point
of black hole evaporation than that employing extremal dilaton black holes.
It is based on the same general principle: in a theory of dynamical
geometry, things can be bigger on the inside than they are on the outside.
We remain convinced that the most plausible scenario for pair production
of such large extended objects is via nucleation of a small geometry
which then expands classically.  As in the inflationary universe, such a
process is not able to access most of the states that can be accomodated
in the eventual large geometry.  Correspondingly, virtual processes will
all involve the temporary appearance of finite volume geometries, which
then subside into the vacuum.  The intermediate set of states will
always be finite (given an ultraviolet cutoff).
These observations reconcile the existence of
essentially infinite repositories of information, with the absence of
observable production amplitudes for these exotic states in ordinary
processes.

In our view, the version of the remnant scenario that we have presented
is the most conservative resolution of the paradox of Hawking radiation
that has been proposed.  The apparently conservative idea that ``all the
information comes out in subtle correlations'' seems to require us to
envisage large quantum corrections to classical geometries in regions in
which the classical description shows no sign of breakdown, while
Hawking's original proposal of nonunitary evolution of density matrices
seems to resist incorporation into a local effective theory of low
energy processes\bps .  In the horned particle scenario, the physics of
dynamical geometry turns out to be stranger than we had imagined, but
not stranger than we are capable of imagining.

\vfill\eject

\centerline{\bf Appendix 1 - Field Equations For the Lagrangian of\lagus}

Given the Lagrangian written in terms of the fields $u$ and $\sigma$, we
can derive the
following Euler-Lagrange equations;

\eqn\ufeq{
2uQ^2e^{-4\sigma} - 2uR + 4u (\partial \sigma)^2 + 8 \partial_{\mu}
(\sqrt{-g}g^{\mu\nu}\partial_\nu u) - 4 u e^{-2 \sigma} = 0
}

\eqn\sigfeq{
u^2 e^{-2 \sigma} - \partial_{\mu}(u^2 \sqrt{-g} g^{\mu\nu}
\partial_{\nu} \sigma) - u^2 Q^2 e^{-4 \sigma} = 0
}

\eqn\tressfeq{
\eqalign{&-(\nabla_\mu \nabla_\nu - g_{\mu\nu} \nabla^2) u^2 - 2 u^2
\partial_\mu \sigma \partial_\nu \sigma + 4 \partial_\mu u
\partial_\nu u \cr& - g_{\mu\nu}(u^2 e^{-2 \sigma} + 2(\partial u)^2 -
u^2(\partial \sigma)^2  - {1 \over 2} Q^2 u^2 e^{-4\sigma}) = 0}
}

and in particular;

\eqn\too{
T_{00} = g_{00}g^{11} \nabla_1^2 u^2 - 2 u^2 \dot \sigma^2 + 4 \dot u^2 -
g_{00}(u^2e^{-2\sigma} + 2(\partial u)^2 -
 u^2 (\partial \sigma)^2 - {1 \over 2} Q^2 u^2 e^{-4\sigma}) = 0
}
\vfill\eject

\centerline{\bf Appendix 2. - The Matching Equations}

To carry out the expansion in powers of $n$ towards the interior of
the
shell we will express all functions in terms of $\tau$ and $n$. In
particular, we will need the derivative in the direction
orthogonal to the shell, and so need ${\partial \over {\partial n}}$
in terms of ${\partial \over {\partial t}}$ and
${\partial \over {\partial r}}$.

Consider the general co-ordinate transformation given by $(\tau,n)
\rightarrow (t(\tau,n),r(\tau,n))$. Using the known forms of the
metric inside and outside the shell we find the following equations.

\eqn\metI{
\eqalign{- 1 &= g_{\tau \tau}\cr & = g_{t t} ({\partial t \over {\partial
\tau}})^2 +
g_{r r} ({\partial r \over {\partial \tau}})^2\cr & =
- ({\partial t \over {\partial \tau}})^2 + {1 \over {(1 - {Q \over
r})^2}} ({\partial r \over {\partial \tau}})^2}
}

and

\eqn\metII{
\eqalign{1 &= g_{n n}\cr & = - ({\partial t \over {\partial n}})^2 + {1 \over
{(1 - {Q \over
r})^2}} ({\partial r \over {\partial n}})^2}
}

and

\eqn\metIII{
\eqalign{0 &=g_{\tau n}\cr&= - {\partial t \over {\partial n}}{\partial t \over
{\partial \tau}}
+ {1 \over {(1 - {Q \over r})^2}}({\partial r \over {\partial
n}})({\partial r \over {\partial \tau}})}
}

Now put $({\partial r \over {\partial \tau}}) \vert_{n=0} = \dot R(\tau)$.

Then \metI gives

\eqn\resI{
{\partial t \over {\partial \tau}} = \sqrt{1 + {1 \over {(1 - {Q \over
R})^2}} \dot R(\tau)^2}
}

and using \metII and \metIII

\eqn\resII{
{\partial r \over {\partial n}} = \sqrt{\dot R(\tau)^2  + {(1 - {Q
\over R})^2}}
}

For a thin shell of matter that acts as a source for the stress tensor
of the theory with Lagrangian \lagmatt,  we have

\eqn\matchI{
\mu(\tau) = \int_{-\epsilon}^{\epsilon} T_{00} dn
}

where $\mu(\tau)$ is the energy density of the shell in the $(\tau,n)$
co-ordinates. Only the singular part of $T_{00}$ will contribute to
\matchI and by looking at the expression for $T_{00}$, we see that (as
a consequence of the continuity of $u$ and $\sigma$ across the
boundary of the shell);

\eqn\matchII{
M u(\tau,0)^2 = \int_{-\epsilon}^{\epsilon} g_{00} g^{11} \nabla^2(u^2)
dn = 2 u \partial_n u \vert_{-\epsilon}^{\epsilon}
}

Using the results of Appendix 1. we find that at $n = -\epsilon$;
\eqn\matchIII{
u \partial_n u = f_1(\tau) R(\tau)^2 e^{-2\phi_0}(1 - {Q \over R})
}
and at $n = \epsilon$;
\eqn\matchIV{
u \partial_n u = e^{-2\phi_0} R (1 - {Q \over {2R}}) \sqrt{\dot
R(\tau)^2  + {(1 - {Q \over R})^2}}
}

So we have the final form of the matching condition;
\eqn\matchV{
M R^2(1 - {Q \over R}) = R (1 - {Q \over {2R}}) \sqrt{\dot
R(\tau)^2  + {(1 - {Q \over R})^2}} + R^2 f_1 (1 - {Q \over R})
}

Finally, we note that the parameter $M$ in the above equations is the
ADM mass of the extremal dilaton black hole:$M = {Q e^{-\phi_0} \over 2}$.

\vfill\eject

\centerline{\bf Appendix 3 - An Ansatz for the Motion of the Shell}

As in the text we assume that $f_{1}(\tau) = {a \over R(\tau)}$, which
gives an ordinary differential equation
for $R(\tau)$. Let's first analyse this equation for
$\tau \rightarrow \infty$, by assuming that $R(\tau) = Q +
\epsilon(\tau)$ with $\epsilon << Q$. Then $\dot \epsilon = -\gamma
\epsilon$, and thus $R(\tau)$ behaves for $\tau \rightarrow \infty$ in
the manner asserted in the text.

Let's now plug the ansatz into the matching equation and solve for
$\dot R(\tau)$. This gives,

\eqn\ode{
\eqalign{
\dot R(\tau) &= - {(R - Q) \over {R(2 R - Q)}}\sqrt{4(f_1R^2 - M
R^2)^2 - (2 R - Q)^2} \cr &=- {(R - Q) \over {R(2 R - Q)}}\sqrt{4(a R - M
R^2)^2 - (2 R - Q)^2
}}
}

This can be integrated numerically to obtain the solution for a given
choice of $R(0)$ and $\dot R(0)$.

For $\dot R(0) = 0$ at $R(0) = 5$, and with a = 109/10, we find the
solution displayed in Fig 4. - a smooth collapse from $R = 5$ to $R=1$.
If we then look at the remaining equations, expanded to the lowest order
in $n$, then we have five equations for the five functions,
$d_1(\tau), h_1(\tau), g_1(\tau), d_2(\tau)$ and $f_2(\tau)$, with all
coefficients going to zero or a finite number as $\tau \rightarrow
\infty$. So this provides a regular collapse situation for this
ansatz.  We could now proceed to solve for higher order terms in the
power series expansion around the shell, finding smooth solutions for
all $\tau$.  Of course, this procedure does not guarantee a solution
which is everywhere smooth.

To analyse the collapse situation completely, one should first find a
time dependent solution to the field equations, (ignoring for the
moment the matching equations) arising from some
smooth initial data on a Cauchy hypersurface, which has a dilaton
 field that is an increasing function
of time which approaches infinity asymptotically. One then uses the
matching equations to join this solution onto
the GMGHS solution along the collapsing shell.
As indicated in the text, we believe that there is a great deal of
freedom in this procedure, and we have not yet been able
to find a simple ansatz which gives explicit solutions for the interior
of the shell.

\vfill\eject
\centerline{\bf Acknowledgements}

We would like to thank our colleagues at Rutgers, N.Seiberg,
 M.Douglas, J.Shapiro,
and especially A.Dabolkhar and S. Shenker
for numerous useful comments and discussions.
T.B. would like to thank the participants in the Aspen Workshop on
Quantum Black Holes, and in particular
A. Strominger and J.Preskill for asking a
number of hard questions that led us to greater understanding.
Suggestions by P. Ginsparg on the subject of titles are warmly acknowledged.
Finally, we would like to thank the referee of the original version of
this paper for the following report: {\it This paper is almost certainly
wrong: production rates for ``Cornucopia'' should be determined by their
masses, and should not be suppressed by volume.   However, publishing it
will offer others exciting and entertaining opportunities to find the
flaws in the authors' logic.  Furthermore, the paper is certainly
interesting and will stimulate further debate on the notorious black
hole information paradox.  It should therefore be published and allowed
to perish.}  These words stimulated us to find the flaws in our own
logic, and we hope that as a consequence our work will not suffer the
fate that the referee envisaged for it.
This work was supported in part by a grant from
the Department of Energy, Number DE-FG05-90ER40559.

\vfill\eject
\centerline{\bf Figure Captions}

\noindent
{\bf Figure 1.} - The Static Cornucopion Geometry at Fixed Polar Angle

\noindent
{\bf Figure 2.} - Instantaneous Snapshot of a Collapsing Cornucopion

\noindent
{\bf Figure 3.} - A Pictorial Argument for Crossing Symmetry

\noindent
{\bf Figure 4.} - Trajectory of the Collapsing Shell

\noindent
{\bf Figure 5.} - Time Slices of an Euclidean Trajectory for Pair
Producing Cornucopions

\listrefs
\end